\documentclass[epj,showpacs]{revtex4}
\usepackage{amsfonts}
\usepackage{amsmath}
\usepackage{graphicx}

\usepackage{amssymb}

\begin{document}

\title{The first passage problem for diffusion through a cylindrical pore with sticky walls\ }
\author{Nicholas A. Licata} 
\email{licata@mpi-cbg.de}
\author{Stephan W. Grill}
\email{grill@mpi-cbg.de}
\affiliation{Max Planck Institute for the Physics of Complex Systems, N\"{o}thnitzer Stra\ss e 38, 01187 Dresden, Germany  \\ and \\
Max Planck Institute for Molecular Cell Biology and Genetics, Pfotenhauer Stra\ss e 108, 01307 Dresden, Germany}

\begin{abstract}
We calculate the first passage time distribution for diffusion through a cylindrical pore with sticky walls. \ A particle diffusively explores the interior of the pore through a series of binding and unbinding events with the cylinder wall. \  Through a diagrammatic expansion we obtain first passage time statistics for the particle's exit from the pore.  Connections between the model and nucleocytoplasmic transport in cells are discussed. 
\end{abstract}
\pacs{87.16.dp, 87.16.ad, 87.15.Vv}

\maketitle

\section{Introduction}

A current topic of considerable interest in cell biology is determining how cells regulate transport of material between the cytoplasm and nucleus \cite{NPCreview,dwelltime,importtime,Gorlich,Imaging}.  This question is intimately related the study of diffusion in confined geometries through the nuclear pore complex (NPC), a roughly cylindrical channel which plays the role of gatekeeper to the nucleus.  Thousands of NPC's are located on the nuclear envelope, and all material which enters or exits the nucleus must pass through individual NPC's.  Many of the proteins which make up the NPC are known to contain phenylalanine-glycine (FG) repeats, which play an important role in the transport process \cite{FG,gating,nucleoporin}.  Namely, cargo destined for the nucleus form a complex with an importin protein which binds specifically to the FG repeats \cite{Importin}.  A microscopic model which seeks to describe transport through the NPC should incorporate both the effects of geometrical confinement and interactions between the importin-cargo complex and FG repeats lining the NPC.  This is in contrast to previous studies, which consider the problem within the framework of one dimensional diffusion in an effective potential \cite{NPCbarrier,NPCtheory}.  

\begin{figure}[t]
\includegraphics[height=3.5009in]{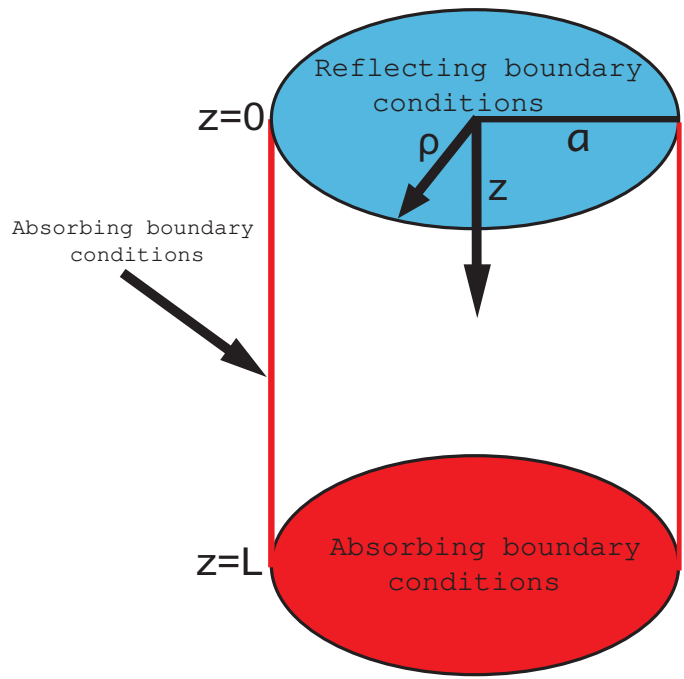}
\caption{(Color online). The geometry and boundary conditions for the first passage problem.  The cylinder has radius $\rho=a$ and length $L$ along the axial 
direction $z$.  A particle enters the cylinder at $z=0$ and is absorbed at the end cap located at $z=L$.  Particles that are absorbed at the side wall will be re-injected into the cylinder (see Fig. \ref{hitexpansion}), where they resume diffusing until encountering the end cap.}
\label{cylinder}
\end{figure}

In this paper we calculate the first passage time statistics for a particle diffusing in a cylindrical pore with sticky walls.  The model is motivated by the nucleocytoplasmic transport problem, however we will make a number of simplifying assumptions.  The first passage problem is formulated in a generic manner which incorporates the effects of diffusion in a confined geometry with an attractive interaction at the system boundary.  We restrict our attention to a cylindrical geometry.  Consider a cylinder of radius $a$ and length $L$, with the axis of the cylinder along the $\hat{z}$ direction (see Fig. \ref{cylinder}).  A particle enters the cylinder at $z=0$.  What is the mean residence time for the particle inside the cylinder?  What is the distribution of exit times through the cylinder cap located at $z=L$?  

There are two dimensionless parameters in the problem, the aspect ratio $\ell=\frac{L}{a}$, and $y=\frac{Dt_{b}}{a^{2}}$, which is a ratio of the characteristic time the particle stays bound to the wall $t_{b}$ to the time required for the particle to diffuse from the center of the cylinder to the wall $\frac{a^{2}}{D}$.  Here $D$ is the diffusion coefficient of the particle.  For a cylinder of length $L$, one could write the mean first passage time through the cylinder in the scaling form $\langle \tau_{c} \rangle = \frac{L^{2}}{2D}g(\ell,y)$.  In the limit of small $\ell$ we have $g(\ell \rightarrow 0,y) =1$, since the particle will always reach the end cap before encountering the wall and the binding is not important. \ In the opposite limit of large $\ell$ the diffusion is effectively one dimensional.  If the binding time becomes negligible, we have $g(\ell \rightarrow \infty,y \rightarrow 0)=1$.  \ We will not pursue a more detailed scaling analysis here.  \ Instead, we will demonstrate how the first passage problem can be formulated to treat the case of an arbitrary distribution function for the binding time $t_{b}$ at finite aspect ratio $\ell$.  To do so, consider drawing all the possible paths of the particle through the cylinder.  We organize these paths by the number of encounters with the cylinder wall.  The simplest paths are those for which the particle diffuses to the end of the cylinder without encountering the wall.  The next set of paths are for particles which diffuse, stick to the wall for some time, detach and diffuse to the end of the cylinder.  Due to the absorbing boundary conditions on the cylinder wall, the detachment process is modeled by radially re-injecting the particle into the cylinder, at which point it resumes diffusing (see Fig. \ref{hitexpansion}).  As a result, the mean first passage time will depend not only on the distribution of residence times at the cylinder wall, but also on the distribution for the particle's radial re-injection.  A detailed discussion of this matter is presented in section 4.  We restrict our attention to the case where the particle cannot diffuse along the surface of the wall at $\rho=a$.  This restriction neglects a class of intermittent problems \cite{Intermittent}, where diffusion in the bulk is coupled to diffusion along the surface of the cylinder.  More complicated paths can be constructed in the same fashion, with more than one encounter with the cylinder wall.

The structure for the paper is the following.  In section 2 we solve the diffusion equation inside the cylinder (see Fig. \ref{cylinder}), with reflecting boundary conditions at $z=0$, and absorbing boundary conditions at $z=L$ and $\rho=a$.  The reflecting boundary condition at $z=0$ means that the particle current vanishes on this surface, and all particles eventually transit the pore.  The choice of absorbing boundary conditions on both the top and bottom surfaces is treated in an appendix.  The main result is a Fourier-Bessel expansion for the Laplace transform of the Green's function (see Eq. \eqref{Green}), from which the particle current and first passage distribution can be determined \cite{Redner}.  In section 3 we calculate the first passage distributions to the wall and end cap.  As a result we obtain the conditional mean first passage time (see Eq. \eqref{tau}) for particles which traverse the pore without encountering the wall at $\rho=a$.  In section 4 the formalism is extended to include interactions between the particle and the wall.  We solve a recursion relation which gives the corrections to the first passage distribution from particle trajectories which include multiple binding and unbinding events with the cylinder wall.  In section 5 we perform an inverse Laplace transform to determine the first passage distribution in the time domain.  We discuss how moments of the first passage distribution can be obtained for the higher order corrections.  
\section{Diffusion Equation}

Consider the defining relation for the Green's function of the diffusion equation, 
\begin{equation}
\left(  \frac{\partial}{\partial \tau} - D \bigtriangledown ^{2} \right) G({ \bf r }, \tau | { \bf r'} ) = \delta(\tau) \delta( {\bf r} - {\bf r' }).
\end{equation}
Here $\tau=t-t'$ is the difference between the observation time $t$ and the source time $t'$, and $D$ is the diffusion coefficient for the particle.  \  Rescaling all lengths by $L$, all times by $L^{2}/D$, and performing a Laplace transform with respect to the time variable $\tau$ we have
\begin{equation}
\label{dif}
\left( s - \bigtriangledown ^{2} \right) \hat{G}({ \bf r }, s | { \bf r'} ) = \delta( {\bf r} - {\bf r' }). 
\end{equation}
Here the Laplace transform of the Green's function is defined as
\begin{equation}
\hat{G}({ \bf r }, s | { \bf r'} )=\int_{0}^{\infty}d\tau\exp(-s\tau) G({ \bf r }, \tau | { \bf r'} ).
\end{equation}
We choose to work in cylindrical coordinates ${ \bf r } = (\rho,z,\phi)$, so the Laplacian operator takes the following form 
\begin{equation}
\bigtriangledown ^{2}\hat{G} = \frac{1}{\rho} \frac{\partial}{\partial \rho} \left( \rho \frac{\partial \hat{G}}{\partial \rho} \right) + \frac{1}{\rho^{2}} \frac{\partial^{2} \hat{G}}{\partial \phi^{2}} + \frac{\partial^{2} \hat{G}}{\partial z^{2}}.
\end{equation}
In the rescaled coordinates, the ends of the cylinder are located at $z=0$ and $z=1$, and the cylinder wall is located at $\rho=1/\ell$ where the dimensionless length scale $\ell=L/a$.  Eq. \eqref{dif} can be solved by a Fourier-Bessel expansion \cite{Jackson,Watson} 
\begin{equation}
\hat{G}({ \bf r }, s | { \bf r'} ) = \frac{1}{2 \pi} \sum_{m=- \infty}^{\infty} \exp(im(\phi - \phi')) \sum_{n=1}^{\infty} J_{m}(x_{mn}\ell\rho) \, \hat{Z}(z,z').
\label{bessel}
\end{equation}
Here $J_{m}(x)$ is the Bessel function of the first kind of order $m$, and $x_{mn}$ is the $n^{th}$ root of the equation $J_{m}(x)=0$.  \  Inserting Eq. \eqref{bessel} into Eq. \eqref{dif} yields the following equation for the unknown function $\hat{Z}(z,z'),$
\begin{equation}
\frac{d^2 \hat{Z}}{dz^2} - \omega_{mn}^{2} \hat{Z}  = -A_{mn}(\rho') \delta(z-z').
\label{zfunc}
\end{equation}
Above the eigenvalue $\omega_{mn}=\sqrt{s+(x_{mn}\ell)^{2}}$ , and the coefficient $A_{mn}(\rho')$ is determined from the Bessel expansion of the radial delta function,
\begin{equation}
A_{mn}(\rho') = \frac{2\ell^2}{J_{m+1}^{2}(x_{mn})} \int_{0}^{1/\ell}  \frac{\delta(\rho - \rho') }{\rho} \,  J_{m}(x_{mn}\ell\rho) \, \rho  \, \mathrm{d} \rho = \frac{2\ell^2}{J_{m+1}^{2}(x_{mn})} \, J_{m}(x_{mn}\ell\rho').
\end{equation}
The solution with the proper symmetry which satisfies the reflecting boundary condition $\frac{d \hat{Z}}{dz} = 0$ at $z=0$, the absorbing boundary condition $\hat{Z}=0$ at $z=1$, and has the proper discontinuous first derivative is 
\begin{equation}
\label{Zeqn}
\hat{Z}(z,z') = \frac{A_{mn}(\rho')\,\text{sech}(\omega_{mn})}{\omega_{mn}}\cosh(\omega_{mn}z_{<})\sinh(\omega_{mn}(1-z_{>})).
\end{equation}
Here $z_{<}$, $z_{>}$ is the smaller, larger of $z$ and $z'$.  \  We arrive at the following expansion for the Green's function
\begin{eqnarray}
\label{Green}
\hat{G}({ \bf r }, s | { \bf r'} ) = \frac{1}{2 \pi} \sum_{m=- \infty}^{\infty} \exp(im(\phi - \phi')) \sum_{n=1}^{\infty} \frac{2\ell^2}{J_{m+1}^{2}(x_{mn})} J_{m}(x_{mn}\ell\rho) J_{m}(x_{mn}\ell\rho') \times \\ \frac{\text{sech}(\omega_{mn})}{\omega_{mn}} \cosh(\omega_{mn}z_{<})\sinh(\omega_{mn}(1-z_{>})). \notag
\end{eqnarray}

With this result, we are in a position to determine the first passage distributions to the wall and end cap.  
\section{First Passage Distributions}

From the Green's function a number of quantities of interest can be calculated. \  In particular, we require the first passage distributions to the wall of the cylinder at $\rho=1/l$ and to the end cap at $z=1$.  \  The current $\hat{\jmath}_{c}(s | { \bf r'})$ through the end cap can be calculated from the Green's function as
\begin{eqnarray}
\hat{\jmath}_{c}(s | { \bf r'})&=& - \int_{0}^{2 \pi} \mathrm{d}\phi \int_{0}^{1/\ell}  \mathrm{d}\rho \,  \rho\, \left. \frac{\partial \hat{G}( { \bf r }, s | { \bf r' })}{\partial z} \right|_{z=1} \\
&=& 2 \sum_{n=1}^{\infty} \frac{\text{sech}(\omega_{0n})}{x_{0n}J_{1}(x_{0n})} J_{0}(x_{0n}l\rho')\cosh(\omega_{0n}z')
\end{eqnarray}
Since the Green's function is normalized, the current $\hat{\jmath}_{c}(s | { \bf r'})$ is the first passage distribution to the end cap for a particle which starts at ${ \bf r' }$.  \  The current through the wall at $\rho=1/\ell$ at height $z$ and hence the first passage distribution can be calculated as
\begin{eqnarray}
\hat{\jmath}_{w}(z, s | { \bf r'})&=& - \int_{0}^{2 \pi}   \mathrm{d}\phi \, \rho  \, \left. \frac{\partial \hat{G}( { \bf r }, s | { \bf r' })}{\partial \rho} \right|_{\rho=1/\ell} \\
&=& 2\ell^2 \sum_{n=1}^{\infty} \frac{x_{0n} \text{sech}(\omega_{0n})}{J_{1}(x_{0n})\omega_{0n}}J_{0}(x_{0n}\ell\rho')\cosh(\omega_{0n}z_{<}) \sinh(\omega_{0n}(1-z_{>})).
\end{eqnarray}

\begin{figure}[t]
\includegraphics[width=5.0548in,height=3.8009in]{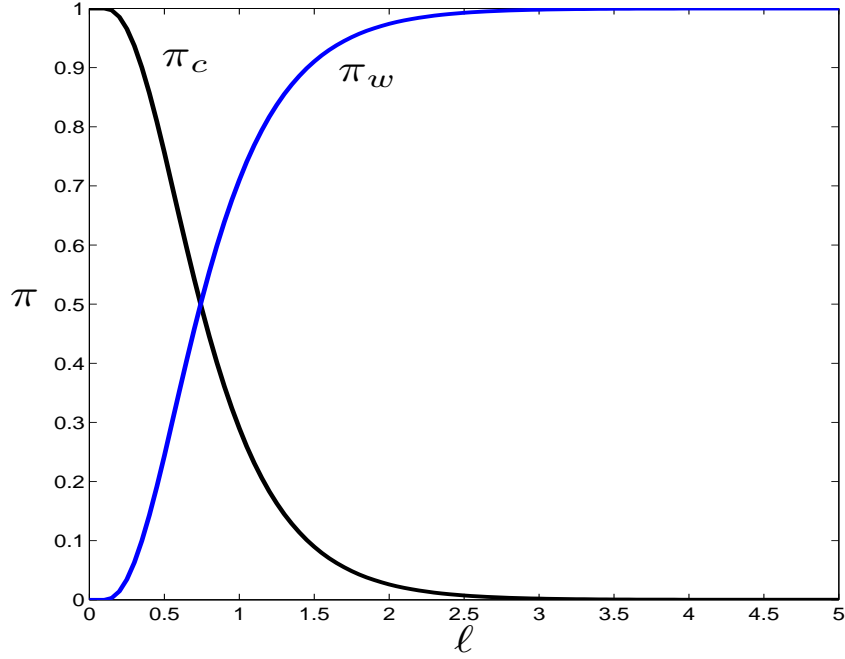}
\caption{(Color online). The splitting probabilities $\pi_{c}$ and $\pi_{w}$ for a particle starting at the origin as a function of the dimensionless aspect ratio $\ell$.  }
\label{splitprob}
\end{figure}

The current $\hat{\jmath}_{w}(z, s | { \bf r'})$ is the first passage distribution for a particle starting at ${ \bf r'} $ to the wall $\rho=1/\ell$ at height $z$.  \  By integrating over $z$ we obtain the first passage distribution to the entire wall, 
\begin{eqnarray}
\hat{\jmath}_{w}(s | { \bf r'}) =  \int_{0}^{1} \mathrm{d}z \, \hat{\jmath}_{w}(z, s | { \bf r'})   = 2\ell^2 \sum_{n=1}^{\infty} \frac{x_{0n} \text{sech}(\omega_{0n})}{J_{1}(x_{0n})\omega_{0n}^{2}}J_{0}(x_{0n}\ell\rho') \times \notag \\
 \{ \sinh(\omega_{0n}(1-z'))\sinh(\omega_{0n}z') - \cosh(\omega_{0n}z')[1-\cosh(\omega_{0n}(1-z'))] \}. 
\end{eqnarray}
These results can be used to obtain the splitting probabilities $\pi_{i}$, where $i \in \{w,c\}$ indexes the wall and end cap respectively.  The splitting probability $\pi_{i}$ is the conditional probability that the particle hits the boundary indexed by $i$.  \  Consider the definition for the Laplace transform of the currents,
\begin{eqnarray}
\hat{\jmath}_{i}(s | { \bf r' }) = \int_{0}^{\infty} \mathrm{d}\tau \, j_{i}(\tau | { \bf r'}) \exp(-s\tau)  &=&  \int_{0}^{\infty} \mathrm{d}\tau \, j_{i}(\tau | { \bf r'})\left[ 1 -s\tau + \frac{s^2\tau^2}{2!} - ... \right]  \\
&=& \pi_{i} \left[ 1 -s \langle \tau \rangle + \frac{s^2\langle \tau^2\rangle}{2!} - ...  \right].
\end{eqnarray}

Hence we see that the splitting probability $\pi_{c}=\hat{\jmath}_{c}(0 | { \bf r' })$ and all the moments of the first passage distribution $\langle \tau_{c}^n \rangle = \frac{(-1)^n}{\pi_{c}} \frac{\partial^{n} \hat{\jmath}_{c}(s | { \bf r' })}{\partial s^{n}} |_{s=0}$ can be obtained by differentiating the Laplace transform of the current.  \  For example, consider a particle starting at the origin $\rho'=z'=0$.  \  The splitting probabilities (see Fig. \ref{splitprob}) $\pi_{c}$ and $\pi_{w}$ to the end cap and wall respectively are (for a particle starting at the origin $\bf{r'=0}$): \
\begin{eqnarray}
\pi_{c} &=& \hat{\jmath}_{c}(0 | {\bf 0})= 2 \sum_{n=1}^{\infty} \frac{\text{sech}(x_{0n}\ell)}{x_{0n}J_{1}(x_{0n})} \\ 
\pi_{w} &=& \hat{\jmath}_{w}(0 | {\bf 0})= 2 \sum_{n=1}^{\infty} \frac{1}{x_{0n}J_{1}(x_{0n})} \left[1 - \text{sech}(x_{0n}\ell) \right]
\end{eqnarray}

\begin{figure}[t]
\includegraphics[width=5.0548in,height=3.8009in]{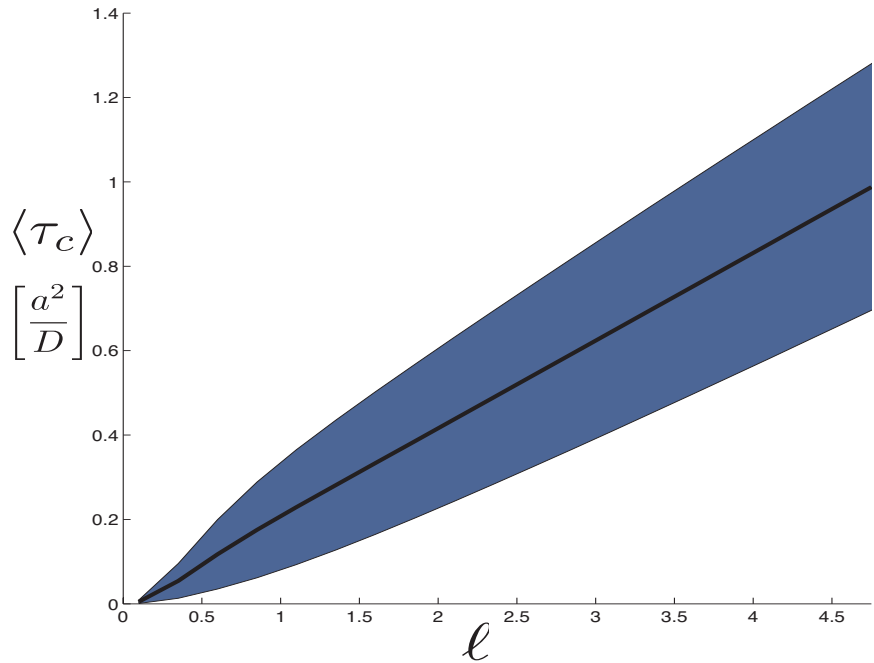}
\caption{(Color online). The conditional mean first passage time $\langle \tau_{c} \rangle$ (see Eq. \eqref{tau}) in units of $\left[ \frac{a^{2}}{D} \right]$ for a particle starting at the origin as a function of the dimensionless aspect ratio $\ell$.  The shaded region covers the range plus and minus one standard deviation $\sigma = \sqrt{\langle \tau_{c}^{2} \rangle - \langle \tau_{c} \rangle ^{2}}$.  }
\label{mfpt}
\end{figure}

These splitting probabilities are propertly normalized as evident from an identity of the zeros of the Bessel function
\begin{equation}
\pi_{c} + \pi_{w} = 2 \sum_{n=1}^{\infty} \frac{1}{x_{0n}J_{1}(x_{0n})} =1.
\end{equation}

Note that the splitting probabilities, which are independent of the angle $\phi$, can also be obtained in a more direct manner by solving Laplace's equation inside the cylinder \cite{Redner}.  For the particle starting at the origin which avoids the wall, the average time required to reach the end
cap is (see Fig. \ref{mfpt})
\begin{equation}
\label{tau}
\langle \tau_{c} \rangle = \frac{1}{\pi_{c}\ell} \sum_{n=1}^{\infty}  \frac{\tanh(x_{0n}\ell)\text{sech}(x_{0n}\ell)}{x_{0n}^{2}J_{1}(x_{0n})}.
\end{equation}

Using this result the limiting behavior can be obtained for the mean first passage time. \ For $\ell \ll 1$ the mean first passage time originally grows quadratically as $\langle\tau_{c}\rangle \simeq  \frac{a^{2}}{D} \frac{\ell^{2}}{2}$, a result we arrived at in the introductory section using scaling arguments.\ For those paths that do not interact with the wall, in the opposite limit of $\ell \gg 1$ we have linear dependence on the aspect ratio as $\langle\tau_{c}\rangle \simeq \frac{a^{2}}{D} \frac{\ell}{2x_{01}}$.  For these two limiting expressions $\langle\tau_{c}\rangle$ is given in dimensionful units to facilitate comparison with Fig. \ref{mfpt}. \ To quantify the width of the distribution the second moment can also be calculated,
\begin{equation}
\langle \tau_{c}^{2} \rangle = \frac{1}{2\pi_{c}\ell^{2}} \sum_{n=1}^{\infty} \frac{\text{sech}(x_{0n}\ell)}{x_{0n}^{3}J_{1}(x_{0n})} \left[  1 + \frac{\text{tanh}(x_{0n}\ell)}{x_{0n}\ell} - 2 \, \text{sech}^{2}(x_{0n}\ell)  \right].
\end{equation}

\section{The hitting number expansion}

\begin{figure}[h]
\includegraphics[width=7in,height=2.5in]{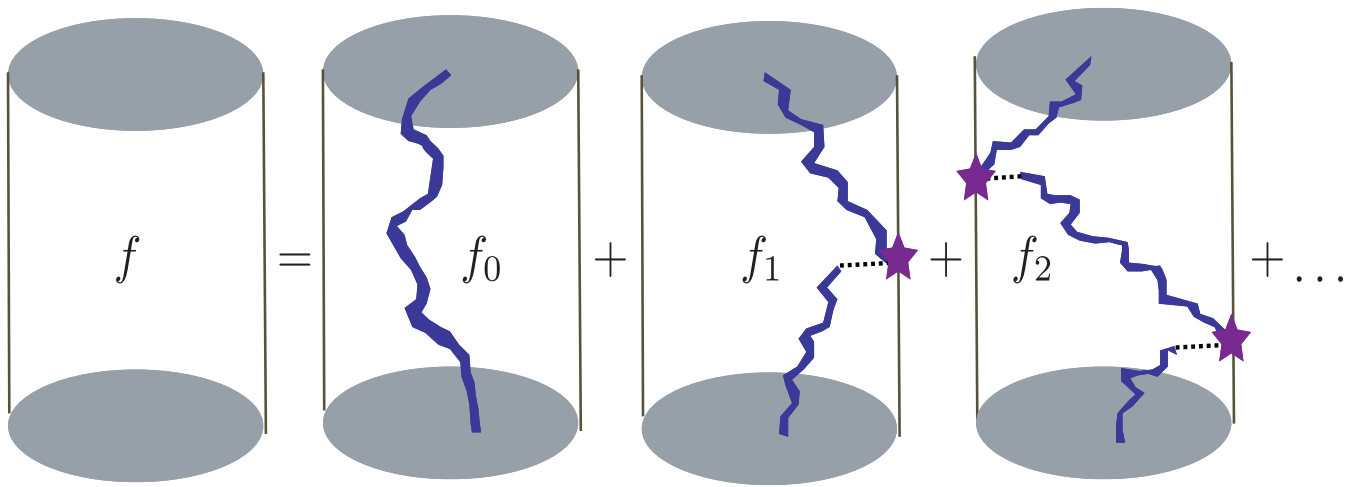}
\caption{(Color online). Graphical depiction of the hitting number expansion.  The particle diffuses until it encounters the absorbing wall.  After some residence 
time at the wall, the particle is radially re-injected into the cylinder, where it resumes diffusing until it exits the cylinder.  }
\label{hitexpansion}
\end{figure}

In this section we will calculate the corrections to these results due to interactions between the particle and the cylinder wall.  We consider a diagrammatic expansion for corrections to the first passage distribution to the end cap (see Fig. \ref{hitexpansion}).  \  As discussed in the introduction, the first such correction is the path for a particle which diffuses, becomes attached to the wall for some time, then detaches and resumes normal diffusion.  \ At this point we should note that the absorbing boundary conditions at the wall present a difficulty for the calculation. \  Because the Green's function vanishes on the wall, the particle cannot resume diffusing there.  In the model, after the absorption phase, the particle is displaced radially inward by a finite amount where it resumes diffusing.  As a result, including the interactions with the wall, the first passage time through the cylinder will depend on two distributions.  The first is the departure time distribution $\Phi(\tau)$, which gives the distribution of residence times at the cylinder wall.  The final result will also depend on the distribution of the starting location for the radial re-injection $P(\rho)$.  Both of these distributions will be discussed in more detail below.  

If we consider all diffusive paths, we can calculate the overall first passage distribution $\hat{f}(s | { \bf r'})$ to the end cap in the following fashion,
\begin{equation}
\hat{f}(s | { \bf r'}) = \sum_{N=0}^{\infty} \hat{f}_{N}(s | { \bf r'}).
\end{equation}

The first term in the series has already been calculated, it is simply the first passage distribution to the wall $\hat{\jmath}_{c}(s | {\bf r'})$.  \ This is a conditional probability distribution, since it provides statistics for particles which avoid the wall during their trip to the end cap, 
\begin{equation}
\hat{f}_{0}(s | { \bf r'}) = \hat{\jmath}_{c}(s | { \bf r'}).
\end{equation}

\begin{figure}[t]
\includegraphics[width=5.0548in,height=3.8009in]{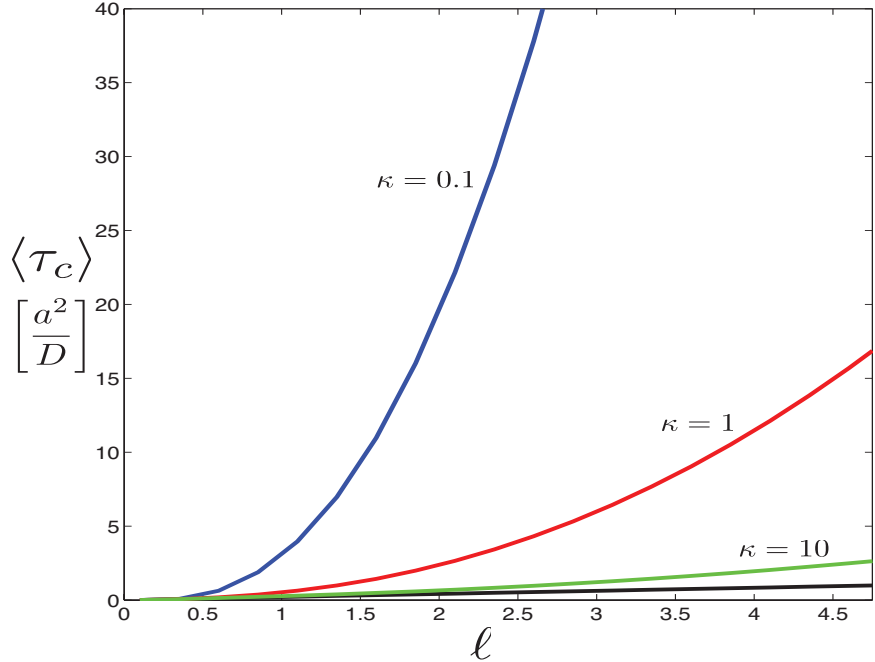}
\caption{(Color online). First order corrections to the mean first passage time $\langle \tau_{c} \rangle$ in units of $\left[ \frac{a^{2}}{D} \right]$ for a particle starting at the origin as a function of the dimensionless aspect ratio $\ell$.  The black curve is the result of Eq. \eqref{tau} which considers only the path $f_{0}$, and the colored curves include the correction $f_{1}$ with different departure rates $\kappa$ (see Eq. \eqref{Arrhenius}).  In the figure, the distribution of starting positions $P(\rho)$ is delta distributed at $\rho_{*}=0.8/\ell$.     }
\label{f1corr}
\end{figure}

The first order correction can be expressed as 
\begin{equation}
\hat{f}_{1}(s | { \bf r'}) = \int _{0}^{1} \mathrm{d}z_{1} \, \hat{\jmath}_{c}(s | z_{1}) \hat{\Phi}( z_{1},s) \hat{\jmath}_{w}( z_{1}, s | { \bf r'}).
\end{equation}
The second order correction is 
\begin{eqnarray}
\hat{f}_{2}(s | { \bf r'}) = \int _{0}^{1} \mathrm{d}z_{1} \, \hat{f_{1}}(s | z_{1}) \hat{\Phi}( z_{1},s) \hat{\jmath}_{w}( z_{1}, s | { \bf r'})= \notag \\
\int _{0}^{1} \mathrm{d}z_{1} \, \int _{0}^{1} \mathrm{d}z_{2} \, \hat{\jmath}_{c}(s | z_{2}) \hat{\Phi}( z_{2},s) \hat{\jmath}_{w}( z_{2}, s | z_{1}) \hat{\Phi}( z_{1},  s) \hat{\jmath}_{w}( z_{1}, s | { \bf r'}).  
\end{eqnarray}
Here we have introduced the Laplace transform $\hat{\Phi}( z , s)$ of the departure time distribution \cite{keylock}.  \ The quantity $\Phi(z ,\tau)d\tau$ is the probability that the particle attached to the wall at height $z$ departs between $\tau$ and $\tau+d\tau$.  \   In many cases this departure distribution is well described by the exponential form \cite{keylock}, \cite{dispersive} 
\begin{equation}
\label{Arrhenius}
\Phi(\tau) = \kappa \exp(-\kappa \tau).
\end{equation}
This exponential form is used in Fig. \ref{f1corr}, but the general results are valid for any choice of $\Phi$ which is independent of $z$.  The result is (see Fig. \ref{f1corr})
\begin{eqnarray}
\hat{f}_{1}(s | { \bf 0}) = \hat{\Phi}(s) \sum_{n=1}^{\infty} \left( \frac{2\langle J_{0}(x_{0n}\ell\rho_{1})\rangle}{x_{0n}J_{1}(x_{0n})} \right) \sum_{m=1}^{\infty} \left( \frac{2\ell^2x_{0m}}{J_{1}(x_{0m})} \right)
\Omega_{mn} \left[ \text{sech}(\omega_{0n})-\text{sech}(\omega_{0m}) \right].   
\end{eqnarray}
Here we have defined 
\begin{equation}
\Omega_{mn} = \frac{1}{\omega_{0m}^{2}-\omega_{0n}^{2}}=\frac{1}{(x_{0m}\ell)^{2}-(x_{0n}\ell)^{2}}.
\end{equation}

\begin{figure}[t]
\includegraphics[width=5.0548in,height=3.8009in]{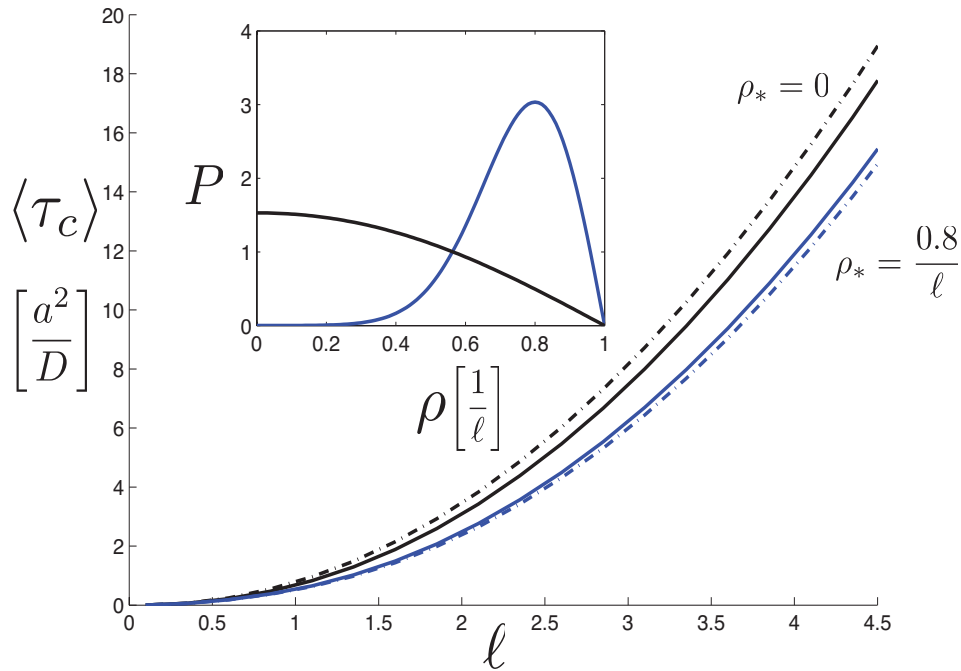}
\caption{(Color online). First order corrections to the mean first passage time $\langle \tau_{c} \rangle$ in units of $\left[ \frac{a^{2}}{D} \right]$ for a particle starting at the origin as a function of the dimensionless aspect ratio $\ell$.  The departure time distribution is given by Eq. \eqref{Arrhenius} with $\kappa=1$.  The black (blue) solid curves are the result with $P(\rho)$ given by Eq. \eqref{weibull} with $k=2$ and $\rho_{*}=0$ $(\rho_{*}= 0.8/\ell)$ respectively.  The corresponding dashed curves are the results with $P(\rho)=\delta(\rho-\rho_{*})/\rho$.  The inset shows the distribution of starting positions $P(\rho)$ as a function of $\rho$ in units of $\left[\frac{1}{\ell}\right]$ as given by Eq. \eqref{weibull} with $\rho_{*}=0.8/\ell$ (blue) and $\rho_{*}=0$ (black).     }
\label{corrmfpt}
\end{figure}

After detaching from the wall, the particle resumes diffusion through the cylinder.  \  Due to the absorbing boundary conditions at $\rho=1/\ell$, the particle cannot resume diffusion at the wall since the Green's function vanishes there. \  For this reason we have averaged over the radial position where the particle begins diffusing after detaching from the wall.  \  If we let $P(\rho)$ denote the distribution function for the starting position, then
\begin{equation}
\langle J_{0}(x_{0n}\ell\rho)\rangle = \frac{\int_{0}^{1/\ell} \mathrm{d}\rho \, \rho \, P(\rho)J_{0}(x_{0n}\ell\rho)}{\int_{0}^{1/\ell} \mathrm{d}\rho \, \rho \, P(\rho)}.
\end{equation}
This model assumption is necessary for the following reasons.  After detaching from the wall, a particle which is displaced a small distance will encounter the wall again with probability one in an arbitrarily short amount of time.  This would result in the particle effectively remaining attached to the wall, and exploring only the surface of the cylinder.  This is contrary to what is observed experimentally in single molecule tracking studies of transport through the NPC \cite{Imaging}.  These studies indicate that the particle spends the majority of its time diffusively exploring the interior of the pore.  For this reason we have introduced the distribution of starting locations $P(\rho)$, which allows us to capture an essential feature of the transport process.  Namely, the particle re-enters the interior of the pore without being immediately re-absorbed, and is able to diffuse in the bulk until it encounters another surface binding domain which is spatially separated from its last encounter.  

We now indicate how the distribution $P(\rho)$ could be calculated.  One possible scenario is that the absorption is modeled as a local potential well at the cylinder wall.  The calculation of $P(\rho)$ can be cast as the problem of determining the exit distribution for the noise activated escape from the potential well, which is a well studied problem in the theory of stochastic differential equations.  For a detailed treatment of this topic for Kramers' escape problem, see \cite{exitdistribution, exitpoint, kramers1, kramers2} and references therein.  Briefly, the particle's Langevin equation is converted into a phase plane system.  The domain of attraction at the bottom of the potential well is bounded by a separatrix, which is the locus of points for which the particle trajectory is equally likely to return to the well or escape.  The exit distribution is then the probability distribution function of the position where escaping trajectories hit the separatrix.  

Alternatively, we consider a scenario for determining the distribution $P(\rho)$ motivated by the structure of the nuclear pore complex (NPC).  In the physical system, the particle binds to FG repeat regions on polymers within the NPC.  In this case $P(\rho)$ is related to the monomer density for the polymers within the NPC.  The monomer distribution for a polymer brush in a cylindrical pore has recently been studied using simulations and scaling theory \cite{cylinderbrush}.   For our present purposes we find that results for the monomer distribution can be approximated using a Weibull like distribution with shape parameter $k$, scale parameter $\lambda$, and normalization constant $C$, 

\begin{equation}
P(\rho) = C \left( \frac{\frac{1}{\ell}-\rho}{\lambda} \right)^{k-1} \exp\left[\left( \frac{\frac{1}{\ell}-\rho}{\lambda} \right)^{k}\right].
\label{weibull}
\end{equation}

In Fig. \ref{corrmfpt} we study the effect of $P(\rho)$ on the first order corrections to $\langle \tau_{c} \rangle$.  We compare the results for a delta distribution $P(\rho)=\delta(\rho-\rho_{*})/\rho$ with a distribution of the form Eq. \eqref{weibull} whith $k=2$ and $\lambda=(\frac{1}{\ell}-\rho_{*})\left(\frac{k}{k-1}\right)^{\frac{1}{k}}$, which has its maximum at $\rho_{*}$.  The qualitative results are very similar.  For a delta distribution, shifting the starting location from near the wall ($\rho_{*}=0.8/\ell$) to the center of the pore ($\rho_{*}=0$) results in a slight increase in $\langle \tau_{c} \rangle$.  As compared to the delta distribution, the distribution given by Eq. \eqref{weibull} results in a small correction.  This correction increases or decreases $\langle \tau_{c} \rangle$ depending on whether walks are weighted more heavily with starting locations $\rho < \rho_{*}$ or $\rho > \rho_{*}$, respectively.  Note that changing $P(\rho)$ effects not only $\langle \tau_{c} \rangle$, but also the probability $\hat{f}_{1}(0 | { \bf 0 })$ for the first order diagram.  This is to be contrasted with the departure time distribution $\Phi(\tau)$, which does not effect the diagram probability.  Changes to the departure time distribution $\Phi(\tau)$ have a more pronounced effect on the first order corrections to the mean first passage time (see Fig. \ref{f1corr}) than changes to the starting distribution $P(\rho)$ (see Fig. \ref{corrmfpt}).  

For a short, fat cylinder, i.e. $\ell \ll 1$, the first few terms in the hitting number expansion suffice to provide accurate results, since the probability of encountering the wall multiple times before reaching the end cap is quite small.  \  We now discuss an iterative procedure to calculate all terms in the hitting number expansion.  \  In principle the calculation can be taken to any desired order.  \ We can obtain any order correction recursively from the relation 
\begin{equation}
\label{recursionrelation}
\hat{f}_{N}(s | { \bf r'}) = \int _{0}^{1} \mathrm{d}z_{1} \, \hat{f}_{N-1}(s | z_{1}) \hat{\Phi}( z_{1},s) \hat{\jmath}_{w}(z_{1},s| { \bf r'}).  
\end{equation}
\  In what follows we assume that the departure time distribution is independent of the spatial coordinate $z$, so that $\hat{\Phi}( z,s)=\hat{\Phi}(s)$.  \ 
In this case the convolution integral can be performed exactly, and we obtain an analytic expression for all orders in the hitting number expansion.  \ These considerations lead to the following expression for the $N^{\text{th}}$ term in the series for a particle which starts at the origin $\bf{r'=0}$, 
\begin{eqnarray}
\label{mainresult}
\hat{f_{N}}(s|{\bf 0}) = \left[  \hat{\Phi}(s) \right]^{N} \sum_{n_{0}=1}^{\infty} \left( \frac{2\ell^{2}x_{0n_{0}}}{J_{1}(x_{0n_{0}})} \right) \prod_{i=1}^{N-1} \sum_{n_{i}=1}^{\infty} \left( \frac{2\ell^{2}x_{0n_{i}} \langle J_{0}(x_{0n_{i}}\ell\rho_{i})\rangle }{J_{1}(x_{0n_{i}})} \right) \times \\ \sum_{n_{N}=1}^{\infty} \left( \frac{2 \langle J_{0}(x_{0n_{N}}\ell\rho_{N})\rangle }{x_{0n_{N}}J_{1}(x_{0n_{N}})} \right) A^{(N)}_{n_{N}n_{N-1}...n_{1}n_{0}}(s). \notag 
\end{eqnarray}
The $N+1$ rank tensor $A^{(N)}(s)$ can be defined recursively as 
\begin{eqnarray}
\label{recursion}
A^{(N)}_{n_{N}n_{N-1}...n_{1}n_{0}}(s) = \Omega_{n_{1}n_{0}}\left[ A^{(N-1)}_{n_{N}n_{N-1}...n_{2}n_{1}}(s) -A^{(N-1)}_{n_{N}n_{N-1}...n_{2}n_{0}}(s) \right].
\end{eqnarray}
Hence all terms in the series are determined after the first is calculated, which is 
\begin{equation}
A^{(1)}_{n_{1}n_{0}}(s) = \Omega_{n_{1}n_{0}} \left[  \text{sech}(\omega_{0n_{0}})-  \text{sech}(\omega_{0n_{1}})  \right].
\end{equation}
The terms with $n_{1}=n_{0}$ must be treated carefully, since direct evaluation gives an indeterminate form. \  Using l'H{\^o}pital's rule we find 
\begin{equation}
A^{(1)}_{nn}(s) = \frac{\text{sech}(\omega_{0n})\text{tanh}(\omega_{0n})}{2\omega_{0n}}.
\end{equation}
Using these results, the second order corrections can be obtained directly from the recursion relation Eq. \eqref{recursion}.  The diagonal term must be treated by the limiting procedure described above which gives
\begin{equation}
A^{(2)}_{nnn}(s)=\frac{\text{sech}^{3}(\omega_{0n})}{16\omega_{0n}^{3}}\left( \omega_{0n}(3-\cosh(2\omega_{0n}))-\sinh(2\omega_{0n}) \right).
\end{equation}

We now have an expression for the Laplace transform of the first passage distribution.  In the next section we illustrate the procedure for obtaining the first passage distribution back in the time domain, and discuss how the moments of the distribution can be obtained by simple differentiation.  
\section{The inverse transform and moments of the distribution}

We now have a result for the first passage distribution in Laplace space.  \  Ideally we would like the result back in the time domain, which requires taking an inverse Laplace transform, 

\begin{equation}
f_{N}(\tau | \text{\bf{0}} ) = \frac{1}{2 \pi i}\int_{\Gamma -i \infty}^{\Gamma + i \infty} \mathrm{d}s \,  e^{s\tau} \hat{f}_{N}(s | \text{\bf{0}}).
\end{equation}

  \  $\Gamma$ is a vertical contour in the complex plane chosen so that all the singularities of $\hat{f}_{N}(s | \text{\bf{0}})$ are to the left of it.  \ We illustrate the procedure with the first diagram, 
  \begin{equation}
  \hat{f}_{0}(s | \text{\bf{0}}) = 2 \sum_{n=1}^{\infty} \frac{\text{sech}(\omega_{0n})}{x_{0n}J_{1}(x_{0n})}.
  \end{equation}

\  To start we note that sech(z) has an infinite number of simple poles at $z=\frac{\pi i}{2}+\pi i k$ with $k \in \mathbb{Z}$. \  To exhibit the poles explicitly we use the infinite series expansion, 
  \begin{equation}
  \text{sech}(z) = \pi \sum_{m=1}^{\infty} \frac{(-1)^{m+1}(2m-1)}{\frac{(2m-1)^{2}\pi^{2}}{4} + z^{2}}.
  \end{equation}
  Hence we first evaluate the integral
  \begin{eqnarray}
  I_{mn} &=& \frac{1}{2 \pi i}\int_{\Gamma -i \infty}^{\Gamma + i \infty} \mathrm{d}s \, \text{sech}(\omega_{0n})  \, e^{s\tau}  \\
  &=& \frac{1}{2  i}  \sum_{m=1}^{\infty} (-1)^{m+1}(2m-1) \int_{\Gamma -i \infty}^{\Gamma + i \infty}\mathrm{d}s \, \frac{e^{s\tau}}{(\alpha_{mn}+s)}. \notag
  \end{eqnarray}
  Here we have defined $\alpha_{mn}=\frac{(2m-1)^{2}\pi^{2}}{4} + (x_{0n}\ell)^{2}$.  The integrand has a pole at $s=-\alpha_{mn}$. \ Since all the poles are on the negative real axis, we can take $\Gamma=0$ and close the contour with a semicircular arc parametrized by $s=Re^{i\phi}$ with $\phi \in [\frac{\pi}{2},\frac{3\pi}{2}]$. \ In the limit $R \rightarrow \infty$ the contribution from the arc vanishes since $e^{st} \sim e^{Rt \text{cos}(\phi)} \rightarrow 0$. \  We evaluate the residue at $s=-\alpha_{mn}$ and obtain 
\begin{eqnarray}
I_{mn} = \pi \sum_{m=1}^{\infty} (-1)^{m+1}(2m-1)e^{-\alpha_{mn}\tau}.
\end{eqnarray}
Using this result we arrive at an expression for the first passage distribution in the time domain, 
\begin{equation}
\label{f0}
f_{0}(\tau | {\bf 0}) = 2 \pi \sum_{n=1}^{\infty} \frac{1}{x_{0n}J_{1}(x_{0n})} \sum_{m=1}^{\infty} (-1)^{m+1}(2m-1)e^{-\alpha_{mn}\tau}.
\end{equation}
So, for the particle which avoids the wall during its journey, the first passage distribution is a sum of exponentials. \  After some transient time, the dominant behavior is exponential with rate constant $\alpha_{11} = \frac{\pi^{2}}{4} + (x_{01}\ell)^{2}$. This is consistent with the roughly exponential distribution of residence times determined experimentally in single molecule tracking studies \cite{dwelltime,Imaging}.  \\
\indent Explicitly performing the inverse transform becomes increasingly difficult for the higher order corrections.  \  As discussed earlier, all the moments of the distribution can be obtained by differentiating with respect to $s$.  \ All of the $s$ dependence in Eq. \eqref{mainresult} is contained in $\hat{\Phi}(s)$ and $A^{(N)}_{n_{N}n_{N-1}...n_{1}n_{0}}(s)$.  
The derivatives of $A^{(N)}_{n_{N}n_{N-1}...n_{1}n_{0}}(s)$ can also be obtained recursively from Eq. \eqref{recursion},  
\begin{equation}
\frac{\partial}{\partial s} A^{(N)}_{n_{N}n_{N-1}...n_{1}n_{0}}(s) = \Omega_{n_{1}n_{0}} \left[ \frac{\partial}{\partial s} A^{(N-1)}_{n_{N}n_{N-1}...n_{2}n_{1}}(s) -\frac{\partial}{\partial s} A^{(N-1)}_{n_{N}n_{N-1}...n_{2}n_{0}}(s) \right].
\end{equation}
The starting point is 
\begin{equation}
\label{derivrecursion}
\frac{\partial}{\partial s} A^{(1)}_{n_{1}n_{0}}(s) = \Omega_{n_{1}n_{0}} \left( A^{(1)}_{n_{1}n_{1}}(s) - A^{(1)}_{n_{0}n_{0}}(s) \right).
\end{equation}
Once the form of the departure time distribution $\Phi$ is specified, we obtain moments of arbitrary order analytically by computing the derivatives of Eq. \eqref{mainresult}.  

\section{Conclusions}
In this paper we calculated the first passage distribution for a particle diffusing through a cylindrical pore with sticky walls.  We organized the calculation in a diagram series which groups diffusive trajectories by the number of encounters with the cylinder wall.  By solving the recursion relation Eq. \eqref{recursionrelation}, we calculate the contribution to the first passage distribution from each diagram.  The main result is given by Eq. \eqref{mainresult}, which gives the contribution to the first passage distribution in Laplace space from each diagram.  This is valid provided that the departure time distribution $\Phi(\tau)$ is independent of the spatial coordinate $z$.  Once the form of the departure time distribution has been specified, all the moments of the first passage distribution can be calculated using the recursion relation and differentiating (see Eq. \eqref{recursion}).  There is a single dimensionless aspect ratio $\ell=L/a$ which encodes information about the geometry of the problem.  We illustrate the procedure for obtaining the first passage distribution in the time domain by performing the inverse Laplace transform on the first diagram.  Considering only the first diagram in the series, Eq. \eqref{f0} provides the probability distribution function for the conditional mean first passage time to the end cap.  

With respect to nucleocytoplasmic transport problem mentioned in the introduction, several comments are in order.  We have made several assumptions in the formulation of our model.  The first passage distribution is obtained for the case of unbiased diffusion \cite{Imaging}.  Because of the reflecting boundary conditions on the top surface, all particles eventually transit the pore.  In experiments not all of the particles which enter the NPC on the cytoplasmic side will successfully make their way to the nucelar side.  As a result, the first passage times predicted by the model should only be compared with experimental events corresponding to a successful passage through the NPC.  An alternative choice of boundary conditions, for example both top and bottom surfaces absorbing, can be easily treated within the model by making the appropriate changes to the boundary conditions satisfied by Eq. \eqref{Zeqn}.  The calculation proceeds in the same manner as before, although now there will be three splitting probabilities instead of two.  The results of this calculation are given in an appendix.  In the model, the particles bind only to the cylinder wall, which corresponds to the FG repeat regions localized on the NPC wall.  We hope that this work will provide a basis for the analysis of future experimental measurements of the first passage distribution through the nuclear pore complex.

We acknowledge S. H. Yoshimura, F. J\"{u}licher, J. S. Bois, and A. V. Klopper for valuable discussions.  

\bibliographystyle{apsrev}
\bibliography{acompat,firstpassage}

\section{Appendix: Absorbing boundary conditions}

In this appendix we derive several results for a modified set of boundary conditions, namely absorbing boundary conditions on all surfaces.  In the paper we treated the top surface as a reflector.  Here we revisit the first passage problem for a cylinder where both the top and bottom surfaces of the cylinder are absorbing, as well as the cylinder wall.  The logic is the same as in the main text.  The unknown function $\hat{Z}(z,z')$ (see Eq. \eqref{zfunc}) must now satisfy the boundary conditions $\hat{Z}=0$ at $z=0$ and $z=1$.  The solution is 
\begin{equation}
\hat{Z}(z,z') = \frac{A_{mn}(\rho')\text{csch}(\omega_{mn})}{\omega_{mn}}\sinh(\omega_{mn}z_{<})\sinh(\omega_{mn}(1-z_{>})).
\end{equation}
As a result, the expansion for the Green's function now reads
\begin{eqnarray}
\label{Green2}
\hat{G}({ \bf r }, s | { \bf r'} ) = \frac{1}{2 \pi} \sum_{m=- \infty}^{\infty} \exp(im(\phi - \phi')) \sum_{n=1}^{\infty} \frac{2\ell^2}{J_{m+1}^{2}(x_{mn})} J_{m}(x_{mn}\ell\rho) J_{m}(x_{mn}\ell\rho') \times \\ \frac{\text{csch}(\omega_{mn})}{\omega_{mn}} \sinh(\omega_{mn}z_{<})\sinh(\omega_{mn}(1-z_{>})). \notag
\end{eqnarray}
There are now three relevant currents.  The current through the top surface $\hat{\jmath}_{t}(s | { \bf r'}) $ at $z=0$ is 
\begin{eqnarray}
\hat{\jmath}_{t}(s | { \bf r'})&=& 2 \sum_{n=1}^{\infty} \frac{\text{csch}(\omega_{0n})}{x_{0n}J_{1}(x_{0n})} J_{0}(x_{0n}l\rho')\sinh(\omega_{0n}(1-z')).
\end{eqnarray}
The current through the bottom surface $\hat{\jmath}_{b}(s | { \bf r'}) $ at $z=1$ is 
\begin{eqnarray}
\hat{\jmath}_{b}(s | { \bf r'})&=& 2 \sum_{n=1}^{\infty} \frac{\text{csch}(\omega_{0n})}{x_{0n}J_{1}(x_{0n})} J_{0}(x_{0n}l\rho')\sinh(\omega_{0n}z').
\end{eqnarray}
The current through the wall $\hat{\jmath}_{w}(z,s | { \bf r'}) $ at $\rho=1/\ell$ is 
\begin{eqnarray}
\hat{\jmath}_{w}(z,s | { \bf r'})&=& 2 \ell^{2} \sum_{n=1}^{\infty} \frac{x_{0n}\text{csch}(\omega_{0n})}{J_{1}(x_{0n})\omega_{0n}} J_{0}(x_{0n}l\rho')\sinh(\omega_{0n}z_{<})\sinh(\omega_{0n}(1-z_{>})).
\end{eqnarray}
Integrating over the wall as before we obtain
\begin{eqnarray}
\hat{\jmath}_{w}(s | { \bf r'}) =  2\ell^2 \sum_{n=1}^{\infty} \frac{x_{0n} \text{csch}(\omega_{0n})}{J_{1}(x_{0n})\omega_{0n}^{2}}J_{0}(x_{0n}\ell\rho') \times \notag \\
 \{ \sinh(\omega_{0n}(1-z'))[\cosh(\omega_{0n}z')-1] - \sinh(\omega_{0n}z')[1-\cosh(\omega_{0n}(1-z'))] \}. 
\end{eqnarray}
As a result we derive the splitting probabilities to the bottom and top of the cylinder
\begin{eqnarray}
\pi_{b}({\bf{r'}}) &=& \hat{\jmath}_{b}(0 | {\bf r'})= 2 \sum_{n=1}^{\infty} \frac{\text{csch}(x_{0n}\ell)}{x_{0n}J_{1}(x_{0n})} J_{0}(x_{0n}\ell\rho')\sinh(x_{0n}\ell z'),\\ 
\pi_{t}({\bf{r'}}) &=& \hat{\jmath}_{t}(0 | {\bf r'})= 2 \sum_{n=1}^{\infty} \frac{\text{csch}(x_{0n}\ell)}{x_{0n}J_{1}(x_{0n})} J_{0}(x_{0n}\ell\rho')\sinh(x_{0n}\ell (1-z')).
\end{eqnarray}
The splitting probability to the wall is
\begin{eqnarray}
\pi_{w}({\bf{r'}}) = \hat{\jmath}_{w}(0 | {\bf r'})= 2 \sum_{n=1}^{\infty} \frac{\text{csch}(x_{0n}\ell)}{x_{0n}J_{1}(x_{0n})} J_{0}(x_{0n}\ell\rho')  \notag \times \\ \{ \sinh(x_{0n}\ell(1-z'))[\cosh(x_{0n}\ell z')-1] - \sinh(x_{0n}\ell z')[1-\cosh(x_{0n}\ell(1-z'))] \}. 
\end{eqnarray}
One can then average over random initial conditions to obtain the averaged splitting probabilities as follows
\begin{eqnarray}
\overline{\pi_{b}} = \frac{1}{\text{Vol(cylinder)}}\int \mathrm{d}^{3}{\bf r'} \pi_{b}({\bf{r'}}) = \frac{4}{\ell} \sum_{n=1}^{\infty} \frac{\text{csch}(x_{0n}\ell)}{x_{0n}^{3}}(\cosh(x_{0n}\ell)-1).
\end{eqnarray}
For uniform initial conditions as considered above symmetry dictates that $\overline{\pi_{t}}=\overline{\pi_{b}}$.  Averaging over the splitting probability to the wall gives
\begin{eqnarray}
\overline{\pi_{w}} = \frac{4}{\ell} \sum_{n=1}^{\infty} \frac{\text{csch}(x_{0n}\ell)}{x_{0n}^{3}}(2 - 2\cosh(x_{0n}\ell)+x_{0n}\ell\sinh(x_{0n}\ell)).
\end{eqnarray}
The averaged splitting probabilities are properly normalized since 
\begin{eqnarray}
\overline{\pi_{b}}+\overline{\pi_{t}}+\overline{\pi_{w}} = 4 \sum_{n=1}^{\infty}\frac{1}{x_{0n}^{2}} = 1.
\end{eqnarray}
The mean first passage time to the bottom of the cylinder can be calculated as before.  The result is 
\begin{eqnarray}
\langle \tau_{b}({\bf{r'}}) \rangle &=&  \frac{-1}{\pi_{b}({\bf r'})} \left. \frac{\partial \hat{\jmath}_{b}(s | { \bf r' })}{\partial s} \right|_{s=0} \nonumber \\ &=&
 \frac{1}{\pi_{b}({\bf r'})\ell} \sum_{n=1}^{\infty} \frac{\text{csch}(x_{0n}\ell)}{x_{0n}^{2}J_{1}(x_{0n})}J_{0}(x_{0n}\ell \rho') \left(  \text{coth}(x_{0n}\ell)\sinh(x_{0n}\ell z') - z' \cosh(x_{0n}\ell z') \right).  
\end{eqnarray}
\end{document}